\documentclass[twocolumn,showpacs,letterpaper,superscriptaddress,prb]{revtex4}%
\usepackage{amsfonts}
\usepackage{amsmath}
\usepackage{amssymb}
\usepackage{graphicx}
\usepackage[colorlinks,linkcolor=blue,citecolor=blue]{hyperref}
\usepackage{times}%
\providecommand{\U}[1]{\protect\rule{.1in}{.1in}}
\begin{document}
\title{Upstanding Rashba spin in honeycomb lattices: Electrically reversible surface
spin polarization}
\author{Ming-Hao Liu}
\email{mhliu@phys.ntu.edu.tw}
\affiliation{Department of Physics, National Taiwan University, Taipei 10617, Taiwan}
\affiliation{Center for Quantum Science and Engineering (CQSE), National Taiwan University,
Taipei 10617, Taiwan}
\author{Ching-Ray Chang}
\email{crchang@phys.ntu.edu.tw}
\affiliation{Department of Physics, National Taiwan University, Taipei 10617, Taiwan}


\begin{abstract}
The spin-split states subject to Rashba spin-orbit coupling in two-dimensional
systems have long been accepted as pointing inplane and perpendicular to the
corresponding wave vectors. This is in general true for free-electron model,
but exceptions do exist elsewhere. Within the tight-binding model, we unveil
the unusual upstanding behavior of those Rashba spins around $\bar{K}$ and
$\bar{K}^{\prime}$ points in honeycomb lattices. Our calculation (i) explains
the recent experiment of the Tl/Si(111)-$(1\times1)$ surface alloy [K.
Sakamoto \textit{et al}., Phys. Rev. Lett. \textbf{102}, 096805 (2009)], where
abrupt upstanding spin states near $\bar{K}$ are observed, and (ii) predicts
an electrically reversible out-of-plane surface spin polarization.

\end{abstract}
\date{\today}
\maketitle

The honeycomb lattice is one of the three types of two-dimensional regular
tessellation---triangular, square, and hexagonal tilings. In solid-state
physics, the honeycomb lattice is described by two staggered triangular
sublattices. The identity of the comprising atoms of the two sublattices
determines if the honeycomb is monoatomic or diatomic, and their interlayer
distance determines whether the honeycomb is flat or bilayer. Thus graphene
and boron nitride are monoatomic and diatomic flat honeycombs,
respectively,\cite{Grosso00} Bi(111) bilayer surface\cite{Hofmann06} is a
monoatomic bilayer honeycomb, and Tl/Si(111)-$\left(  1\times1\right)  $
surface alloy\cite{Sakamoto09} is a diatomic bilayer honeycomb.

Among these honeycomb lattices, graphene has been under the most intensive
investigation due to its unusual Dirac-like electronic
excitations.\cite{Neto09} When deposited on a substrate, the structural
inversion symmetry perpendicular to the graphene plane is broken, and the
Rashba spin-orbit coupling\cite{Bychkov84} hence emerges. Recent experimental
measurement of Rashba spin splitting in graphene on Ni(111) substrate
with\cite{Varykhalov08} and without\cite{Dedkov08} an intercalated Au
monolayer eventually drew theorists' attention to the electronic structure, in
the presence of Rashba coupling, of graphene,\cite{Rashba09} i.e., near
$\bar{K}$ and $\bar{K}^{\prime}$ in monoatomic flat honeycomb. In other
honeycomb systems, Bi(111) bilayer surface is believed to contain strong
Rashba coupling\cite{Koroteev04} and exhibit interesting spin-Hall
patterns,\cite{Liu07Bi111} and a more recent experimental effort even shows an
unusual Rashba spin behavior at $\bar{K}$ point in Tl/Si(111)-$\left(
1\times1\right)  $ surface alloy.\cite{Sakamoto09}

In this Rapid Communication we present a unified tight-binding description to
understand the Rashba effect in honeycomb lattices. To focus on the major
effect brought by the Rashba coupling, we consider electron hopping up to the
nearest neighbors. Particular attention will be paid to the spin configuration
near $\bar{K}$ and $\bar{K}^{\prime}$ points, which shows an abrupt upstanding
Rashba spin behavior, in good agreement with Ref. \onlinecite{Sakamoto09}.
Moreover, we show that the upstanding spins along $\bar{K}$ and $\bar
{K}^{\prime}$ directions are opposite. Accordingly, we propose an electrically
reversible out-of-plane surface polarization, which will be numerically shown
by Landauer-Keldysh formalism.

Consider a honeycomb lattice constructed by primitive translation vectors
$\mathbf{t}_{1}=a(1/2,\sqrt{3}/2,0)\ $and\ $\mathbf{t}_{2}=a(-1/2,\sqrt
{3}/2,0)$, and basis vectors $\mathbf{d}_{1}=0$ and $\mathbf{d}_{2}%
=a(0,1/\sqrt{3},d_{z})$, where $a$ is the lattice constant. For flat
honeycombs we have $d_{z}=0$, and $d_{z}\neq0$ describes a bilayer case. We
begin with the $4\times4$ tight-binding Hamiltonian\cite{Liu07Bi111} of the
Slater and Koster type,\cite{Grosso00,Slater54}%
\begin{equation}
\mathbb{H}=\left(
\begin{array}
[c]{cc}%
\mathbb{H}_{11} & \mathbb{H}_{12}\\
\mathbb{H}_{12}^{\dag} & \mathbb{H}_{22}%
\end{array}
\right)  , \label{H4x4}%
\end{equation}
with off-diagonal element%
\begin{equation}
\mathbb{H}_{12}=\left(
\begin{array}
[c]{cc}%
U(1+2F) & -it_{R}(1-F-\sqrt{3}G)\\
-it_{R}(1-F+\sqrt{3}G) & U(1+2F)
\end{array}
\right)  , \label{H12}%
\end{equation}
where $t_{R}$ is the Rashba hopping strength, $U\equiv l_{z}^{2}V_{pp\sigma
}+\left(  1-l_{z}^{2}\right)  V_{pp\pi}$, $l_{z}$ being the direction cosine
of nearest neighbors ($l_{z}=0$ for flat and $l_{z}\neq0$ for bilayer), is the
two-center interaction integral involving $p_{z}$ atomic orbitals, and the
compact functions are given by $F\equiv\exp(-i\sqrt{3}k_{y}a/2)\cos\left(
k_{x}a/2\right)  $ and $G\equiv\exp(-i\sqrt{3}k_{y}a/2)\sin\left(
k_{x}a/2\right)  .$ The diagonal elements of Eq. (\ref{H4x4}) are
$\mathbb{H}_{ii}=E_{pi}\mathbb{I}$ with $\mathbb{I}$ the $2\times2$ identity
matrix and $i=1,2$. For monoatomic honeycombs, we have $E_{p1}=E_{p2}=E_{p}$
but for diatomic honeycombs, $E_{p1}\neq E_{p2}$. In the following we consider
$E_{p1}=E_{p2}$; straightforward generalization to the diatomic case will be
shown later.

Adopting the same trick of Rashba,\cite{Rashba09} Eq. (\ref{H4x4}) can be
reduced to a $2\times2$ Hamiltonian,%
\begin{equation}
\mathbb{H}_{A}\left(  E\right)  =E_{p}\mathbb{I}+\frac{\mathbb{H}%
_{12}\mathbb{H}_{12}^{\dag}}{E-E_{p}}\label{HA}%
\end{equation}
for sublattice $A$, which depends explicitly on its eigenvalue $E$. The
Schr\"{o}dinger equation of Hamiltonian (\ref{HA}) is $\mathbb{H}_{A}\left(
E\right)  |\psi_{A}\rangle=E|\psi_{A}\rangle$. By solving the characteristic
equation $\det(\mathbb{H}_{A}\left(  E\right)  -E)=0$ the eigenvalues of Eq.
(\ref{HA}) can be written as%
\begin{equation}%
\begin{array}
[c]{cc}%
E_{\mu\nu}=E_{p}+\mu E_{\nu}, & \mu,\nu=\pm1
\end{array}
\label{Emunu}%
\end{equation}
with%
\begin{equation}
E_{\nu}=\sqrt{\frac{\operatorname{Tr}h+\nu\sqrt{\left(  \operatorname{Tr}%
h\right)  ^{2}-4\det h}}{2}},\label{Enu}%
\end{equation}
where%
\begin{equation}
h=\mathbb{H}_{12}\mathbb{H}_{12}^{\dag}=\left(
\begin{array}
[c]{cc}%
h_{11} & h_{12}\\
h_{21} & h_{22}%
\end{array}
\right)  \label{h}%
\end{equation}
will play an important role in the following derivation. The eigenvectors can
be written as either of
\begin{subequations}
\label{eigvec}%
\begin{align}
|\psi_{A}^{\mu\nu}\rangle &  =\left(  \left\vert h_{12}\right\vert
^{2}+\left\vert E_{\nu}^{2}-h_{11}\right\vert ^{2}\right)  ^{-1/2}\left(
\begin{array}
[c]{c}%
h_{12}\\
E_{\nu}^{2}-h_{11}%
\end{array}
\right)  \label{eigvec 1}\\
|\psi_{A}^{\mu\nu}\rangle &  =\left(  \left\vert E_{\nu}^{2}-h_{22}\right\vert
^{2}+\left\vert h_{21}\right\vert ^{2}\right)  ^{-1/2}\left(
\begin{array}
[c]{c}%
E_{\nu}^{2}-h_{22}\\
h_{21}%
\end{array}
\right)  ,\label{eigvec 2}%
\end{align}
which are independent of $\mu$. Both Eqs. (\ref{eigvec 1}) and (\ref{eigvec 2}%
) are valid for carrying out the spin expectation $\langle\vec{S}%
\rangle=(\hbar/2)\langle\vec{\sigma}\rangle$, $\vec{\sigma}=(\sigma^{x}%
,\sigma^{y},\sigma^{z})$ being the Pauli matrix vector, except at the symmetry
points $\bar{\Gamma}$, $\bar{M}$, $\bar{K}$ and $\bar{K}^{\prime}$.
\begin{figure}[b]
\centering\includegraphics[height=8.5cm]{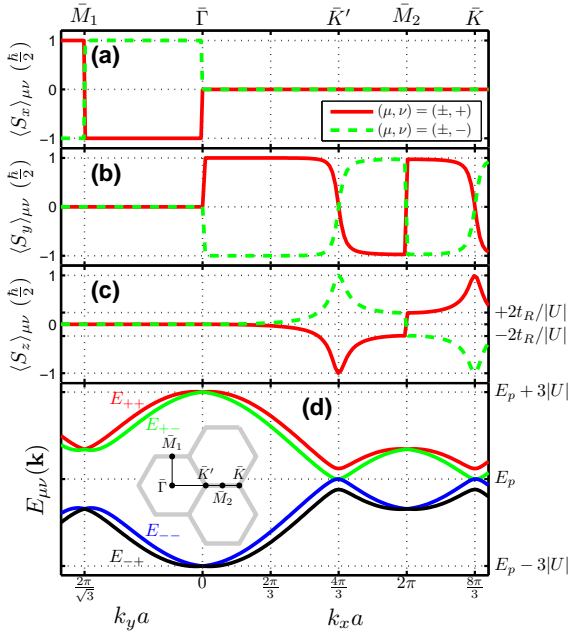}\caption{(Online color)
The spin components (a) $\langle S_{x}\rangle_{\mu\nu},$ (b) $\langle
S_{y}\rangle_{\mu\nu},$ and (c) $\langle S_{z}\rangle_{\mu\nu}$ of energy
eigenstates of a Rashba spin-orbit-coupled monoatomic honeycomb lattice with
its band structure shown in (d). The parameters are set $t_{R}/\left\vert
U\right\vert =0.12$. The Brillouin zone is shown in the inset in (d).}%
\label{FIG1}%
\end{figure}With careful treatment at those points, the spin direction
$\langle\vec{S}\rangle_{\mu\nu}=(\hbar/2)\langle\psi_{A}^{\mu\nu}|\vec{\sigma
}|\psi_{A}^{\mu\nu}\rangle$ based on Eq. (\ref{eigvec})\ subject to the four
eigenstate branches are shown in Figs. \ref{FIG1}(a)--\ref{FIG1}(c); the band
structure according to Eqs. (\ref{H12})--(\ref{h}) is shown in Fig.
\ref{FIG1}(d). Clearly one can see an abrupt upstanding component $\langle
S_{z}\rangle_{\mu\nu}$ near $\bar{K}$ and $\bar{K}^{\prime}$ points. For the
$\nu=+1$ branch, we depict the spin configuration based on Eq. (\ref{eigvec})
in Fig. \ref{FIG2}, where each arrow is determined by $(\langle\sigma
^{x}\rangle_{\mu+},\langle\sigma^{y}\rangle_{\mu+})$ and the color shading is
by $\langle\sigma^{z}\rangle_{\mu+}$. The bright (dark) region of
$\langle\sigma^{z}\rangle\approx1$ ($\langle\sigma^{z}\rangle\approx-1$)
around $\bar{K}$ ($\bar{K}^{\prime}$) can be clearly seen in the main panel of
Fig. \ref{FIG2}, reflecting its inherent $C_{3}$ symmetry.\begin{figure}[t]
\centering\includegraphics[width=7.5cm]{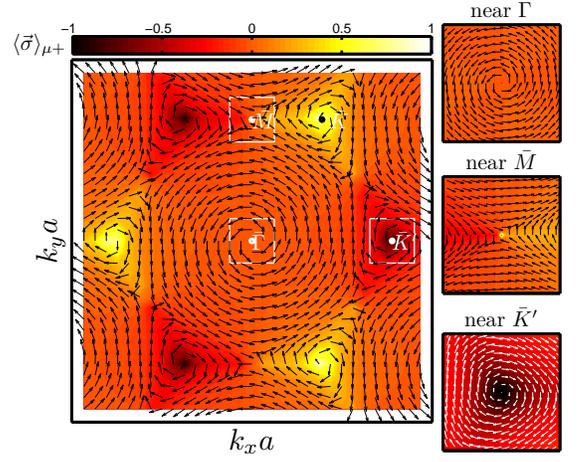}\caption{{}{}(Color
online) Spin configuration for the $\nu=+1$ branche. Each arrow is given by
$(\langle\sigma^{x}\rangle_{\mu+},\langle\sigma^{y}\rangle_{\mu+})$ and the
color shading is determined by $\langle\sigma^{z}\rangle_{\mu+}.$}%
\label{FIG2}%
\end{figure}

To provide deeper understanding of Fig. \ref{FIG1}, as well as the zoomed-in
plots of Fig. \ref{FIG2}, below we give a series of discussion of $E_{\mu\nu}$
and $\langle\vec{\sigma}\rangle_{\mu\nu}$ around those symmetry points.

\paragraph*{The $\bar{\Gamma}$ point.}

Assuming $\mathbf{k}=\vec{\delta}=\delta(\cos\phi,\sin\phi)$ with $\delta
a\ll1$, one can obtain%
\end{subequations}
\begin{equation}
E_{\nu}(\bar{\Gamma}+\vec{\delta})\approx3\left\vert U\right\vert
+\frac{\left(  t_{R}^{2}-2U^{2}\right)  }{8\left\vert U\right\vert }\delta
^{2}a^{2}+\nu\frac{\sqrt{3}}{2}t_{R}\delta a \label{E near G}%
\end{equation}
and $\langle\vec{\sigma}\rangle_{\mu\nu}(\bar{\Gamma}+\vec{\delta})\approx
-\nu(\sin\phi,-\cos\phi,0)$ which is identical to the free-electron case.
Equation (\ref{E near G}) is useful for determining parameters $E_{p},$ $U,$
and $t_{R}$ by matching with the free-electron dispersion $E\left(  k\right)
=E_{0}+\hbar^{2}k^{2}/2m^{\ast}\pm\alpha k.$ The band offset $E_{0}$, the
Rashba parameter $\alpha,$ and the curvature $\hbar^{2}/2m^{\ast}$ (or the
effective mass $m^{\ast}$), as well as the lattice constant $a$ are
experimentally measurable. In the usual $t_{R}^{2}\ll\left\vert U\right\vert
^{2}$ case we have $E_{p}=E_{0}+\left\vert 3U\right\vert ,$ $\left\vert
U\right\vert =2\hbar^{2}/m^{\ast}a^{2},$ and $t_{R}=2\alpha/\sqrt{3}a$.

\paragraph*{The $\bar{M}$ point.}

Let $\bar{M}=\left(  2\pi/a,0\right)  ,$ i.e., $\bar{M}_{2}$ in Fig.
\ref{FIG1}. Assuming $\mathbf{k}=\bar{M}+\vec{\delta}=(2\pi/a+\delta
_{x},\delta_{y})$ one obtains an anisotropically free-electron-like
dispersion,%
\begin{align}
E_{\nu}(\bar{M}+\vec{\delta})  &  \approx\sqrt{U^{2}+4t_{R}^{2}}%
-\frac{\left\vert U\right\vert }{4}\left(  \delta_{x}^{2}-3\delta_{y}%
^{2}\right)  a^{2}\nonumber\\
&  +\nu\frac{\sqrt{3}}{2}t_{R}\sqrt{\delta_{x}^{2}+9\delta_{y}^{2}}a,
\label{E near M}%
\end{align}
which indicates that $\bar{M}$ is a saddle point as one can see from the
different sign of $\delta_{x}^{2}$ and $\delta_{y}^{2}$ in the second term.
This means that the effective mass of the electron at state near $\bar{M}$ has
opposite sign when going \emph{along} and \emph{perpendicular} to the
Brillouin-zone boundary. In addition, the latter has an effective mass 3 times
lighter and an effective Rashba parameter three times stronger than the
former. Equation (\ref{E near M}) therefore explains why we have band shape
near $\bar{\Gamma}$ identical with that near $\bar{M}_{2}$ but not $\bar
{M}_{1}$ [see Fig. \ref{FIG1}(d)].

The anisotropy at $\bar{M}$ also reveals in the corresponding spin direction,%
\begin{equation}
\langle\vec{\sigma}\rangle_{\mu\nu}(\bar{M}+\vec{\delta})\approx\frac{\nu
}{\sqrt{5-4\cos2\phi}}\left(
\begin{array}
[c]{c}%
3\sin\phi\\
\cos\phi\\
\dfrac{2t_{R}}{\left\vert U\right\vert }\cos\phi
\end{array}
\right)  , \label{S near M}%
\end{equation}
where we keep terms up to first order in $t_{R}/|U|$. Clearly from Eq.
(\ref{S near M}) the $z$ component saturates to $2t_{R}/|U|$ along the
Brillouin boundary near $\bar{M}$ but vanishes when going perpendicular to the
boundary [see Fig. \ref{FIG1}(c)].

\paragraph*{The $\bar{K}$ and $\bar{K}^{\prime}$ points.}

Let $\bar{K}^{\prime}=(\mathbf{g}_{1}-\mathbf{g}_{2})/3.$ Assuming
$\mathbf{k}=\bar{K}^{\prime}+\vec{\delta}=\left(  4\pi/3a+\delta_{x}%
,0+\delta_{y}\right)  $ we obtain%
\begin{equation}
E_{\nu}(\bar{K}^{\prime}+\vec{\delta})\approx\frac{1}{2}\sqrt{\left(
3t_{R}\right)  ^{2}+3U^{2}\delta^{2}a^{2}}+\nu\frac{3t_{R}}{2},
\label{E near K'}%
\end{equation}
in agreement with Ref. \onlinecite{Rashba09}. The spin direction near $\bar
{K}^{\prime}$, up to second order in $\left(  \delta a\right)  ,$ is given by%
\begin{equation}
\langle\vec{\sigma}\rangle_{\mu\nu}(\bar{K}^{\prime}+\vec{\delta})=\nu\left(
\begin{array}
[c]{c}%
-\dfrac{U}{\sqrt{3}t_{R}}\delta a\sin\phi\\
\dfrac{U}{\sqrt{3}t_{R}}\delta a\cos\phi\\
-1+\dfrac{1}{6}\dfrac{U^{2}}{t_{R}^{2}}\delta^{2}a^{2}%
\end{array}
\right)  , \label{S near K'}%
\end{equation}
which shows at $\bar{K}^{\prime}$ we have $\langle\sigma^{z}\rangle_{\mu\pm
}\left(  \bar{K}^{\prime}\right)  =\mp1$. Around $\bar{K}$ the dispersion is
identical to Eq. (\ref{E near K'}), and the spin configuration has a reversed
helicity, i.e., opposite out-of-plane component,%
\begin{equation}
\langle\sigma^{z}\rangle_{\mu\nu}(\bar{K}+\vec{\delta})=-\langle\sigma
^{z}\rangle_{\mu\nu}(\bar{K}^{\prime}+\vec{\delta}), \label{SK = -SK'}%
\end{equation}
but unchanged inplane component.

So far the discussion is basically for sublattice $A$ since we have obtained
reduced Hamiltonian (\ref{HA}) by expressing the wave function of sublattice
$B$, $\psi_{B}$, in terms of that of sublattice $A$, $\psi_{A}$. We could have
as well expressed $\psi_{A}$ in terms of $\psi_{B}$; the resulting reduced
Hamiltonian then would be $\mathbb{H}_{B}\left(  E\right)  =E_{p}%
\mathbb{I}+\mathbb{H}_{12}^{\dag}\mathbb{H}_{12}/\left(  E-E_{p}\right)  $,
leading to identical dispersion, identical inplane spin direction, but
opposite $\langle\sigma^{z}\rangle$ component,
\begin{equation}
\langle\psi_{A}|\sigma^{z}|\psi_{A}\rangle=-\langle\psi_{B}|\sigma^{z}%
|\psi_{B}\rangle.\label{SA = -SB}%
\end{equation}
In addition, we have so far focused on the monoatomic honeycomb lattice. For
different sublattice atoms $A$ and $B$, we have $E_{p1}\neq E_{p2}$. The
energy dispersion can be straightforwardly derived as $\varepsilon_{\mu\nu
}(\mathbf{k})=\bar{E}_{p}+\mu\varepsilon_{\nu}$ with $\varepsilon_{\nu}%
=\sqrt{\Delta_{p}^{2}+E_{\nu}^{2}}$, $\bar{E}_{p}=(E_{p1}+E_{p2})/2$,
$\Delta_{p}=(E_{p1}-E_{p2})/2$, and $E_{\nu}$ given by Eq. (\ref{Enu}).
Noticing from Eqs. (\ref{E near G}), (\ref{E near M}), and (\ref{E near K'})
that around the symmetry points $E_{+}\neq E_{-}$ occurs only at $\bar{K}$ and
$\bar{K}^{\prime}$, we expect the modification to the band structure due to
$\Delta_{p}$ appears the most salient at $\bar{K}$ and $\bar{K}^{\prime}$.
Indeed, with $E_{\nu}(\mathbf{k}=\bar{K}$ or $\bar{K}^{\prime})=3t_{R}\left(
1+\nu\right)  /2$ a gap $\varepsilon_{+-}-\varepsilon_{--}=2\varepsilon
_{-}=2\left\vert \Delta_{p}\right\vert $ is opened. At other symmetry points,
$\varepsilon_{\mu\pm}$ remain degenerate. Despite the opened gap, however, the
eigenvectors surprisingly remain the same as Eq. (\ref{eigvec}), and the
upstanding Rashba spin around $\bar{K}$ and $\bar{K}^{\prime}$ described
previously is therefore unchanged.\begin{figure}[t]
\centering \includegraphics[width=8.5cm]{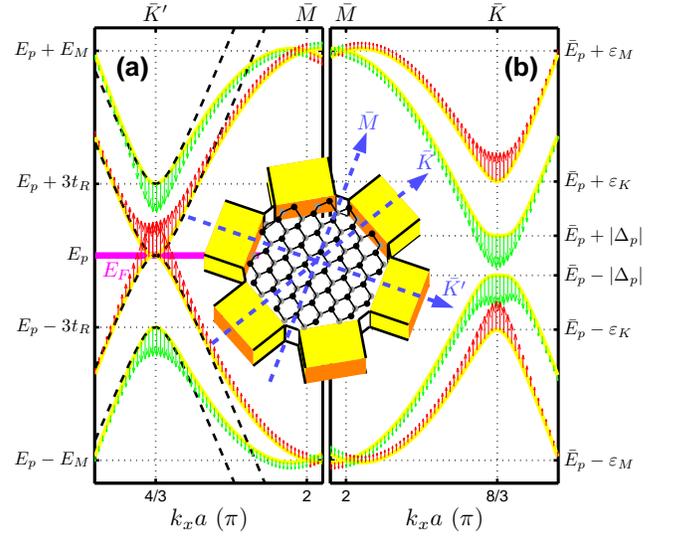} \caption{{}(Color
online) Energy dispersion and $\langle\sigma^{z}\rangle_{\mu\nu}$ (indicated
by arrows) for (a) monoatomic and (b) diatomic honeycomb lattices. In (a) the
black dashed lines are the approximated dispersions given by Eq.
(\ref{E near K'}); $E_{M}=\sqrt{U^{2}+4t_{R}^{2}}$. In (b) we set $\Delta
_{p}=0.1\left\vert U\right\vert $; $\varepsilon_{M}=\sqrt{\Delta_{p}^{2}%
+U^{2}+4t_{R}^{2}}$ and $\varepsilon_{K}=\sqrt{\Delta_{p}^{2}+\left(
3t_{R}\right)  ^{2}}$. The inset shows a six-terminal bilayer honeycomb
channel with black (gray) dots the upper (lower) sublattice.}%
\label{FIG3}%
\end{figure}

To show the opened gap with unchanged spin direction, we plot the energy
dispersion curves with $\langle\sigma^{z}\rangle_{\mu\nu}$ in Fig. \ref{FIG3}
for monoatomic honeycomb near $\bar{K}^{\prime}$ [panel (a)] and diatomic
honeycomb near $\bar{K}$ [panel (b)]. Figure \ref{FIG3}(b) readily explains
the recent spin-resolved angle-resolved photoelectron spectroscopy data for
the Tl/Si(111)-$(1\times1)$ surface alloy.\cite{Sakamoto09} In such surface
alloy,\cite{Lee02,Sakamoto06} the Tl coverage on the Si(111) substrate is one
monolayer, and the topmost Tl (sublattice $A$) and Si (sublattice $B$) layers
form a diatomic honeycomb lattice of the bilayer type ($d_{z}<0$). The
$\mu=+1$ branch in Fig. \ref{FIG3}(b) thus resembles the band feature near
$\bar{K}$ reported in Ref. \onlinecite{Sakamoto09}. It is important to note,
however, that for flat honeycombs such as graphene, both sublattices $A$ and
$B$ will be silmultaneously measured, leading to vanishing $\langle
S_{z}\rangle$, in view of Eq. (\ref{SA = -SB}).

We have shown that the upstanding Rashba spin around $\bar{K}$ and $\bar
{K}^{\prime}$ points is a fundamental property of the spin configuration in
honeycomb lattices, whether flat or bilayer, monoatomic or diatomic. Next we
illustrate how striking this property can be. For simplicity, let us consider
a six-terminal channel made of monoatomic bilayer honeycomb. See the inset in
Fig. \ref{FIG3}. Assume that the transport of this six-terminal device is
supported by the surface states so that electrons are only allowed to hop
within the surface bilayer, i.e., the honeycomb lattice. Let the\ Fermi energy
$E_{F}$ lie just at $E_{p}$. The situation is like Fig. \ref{FIG3}(a).
Recalling Eq. (\ref{SA = -SB}) we expect that when driving the electrons along
$\bar{K}^{\prime}$, the transport states with $E_{F}=E_{p}$ are $+z$ spin
polarized on sublattice $A$ but $-z$ spin polarized on sublattice $B$.
Recalling further Eq. (\ref{SK = -SK'}), the surface spin polarization
(assumed to be contributed by sublattice $A$ only) is out-of-plane and can be
electrically controlled by either $\bar{K}$ or $\bar{K}^{\prime}$ biasing.

To visualize this idea, let us calculate for the six-terminal bilayer
honeycomb channel the local spin densities by employing the
Landauer-Keldysh\cite{Nikolic05,Nikolic06} formalism, subject to Hamiltonian,
$\mathcal{H}=E_{p}\sum_{n}c_{n}^{\dag}c_{n}+\sum_{\left\langle nm\right\rangle
}c_{m}^{\dag}[U\mathbb{I}+it_{R}(\vec{\sigma}\times\mathbf{d}_{mn})_{z}%
]c_{n},$ with $c_{n}^{\dag}$ ($c_{n}$) the creation (annihilation) operator of
the electron on site $n$ and $\mathbf{d}_{mn}$ the unit vector pointing from
site $n$ to $m$. The total number of lattice sites in the honeycomb channel is
480. For clarity we will plot $\langle\sigma^{z}\rangle$ only on the surface
(sublattice $A$). A positive (negative) $\langle\sigma^{z}\rangle$ on each
site will be denoted by a red (green) dot, with the dot size proportional to
the magnitude of $\langle\sigma^{z}\rangle.$ The applied potential energy of
$\pm eV_{0}/2$ will be denoted as \textquotedblleft$\pm$,\textquotedblright%
\ and $eV_{0}=0$ as \textquotedblleft$0$\textquotedblright\ on each lead. The
parameters $E_{p}=0$, $U=-1%
%TCIMACRO{\unit{eV}}%
%BeginExpansion
\operatorname{eV}%
%EndExpansion
$, and $t_{R}/\left\vert U\right\vert =0.12$ are identical to those used in
previous figures, and are within a realistic range.\footnote{The ratio
$t_{R}/\left\vert U\right\vert $ is of the order of $10^{-3}$
(Ref.\ \onlinecite{Varykhalov08}) to $10^{-2}$ (Ref.\ \onlinecite{Dedkov08})
for graphene, and of $10^{-1}$ for Bi(111) bilayer surfaces
(Ref.\ \onlinecite{Koroteev04} and \onlinecite{Liu07Bi111}). The reported
value in Tl/Si(111)-$\left(  1\times1\right)  $ surface alloy
(Ref.\ \onlinecite{Sakamoto09}) leads to an even stronger ratio $t_{R}%
/\left\vert U\right\vert \sim0.4$.} The bias is $eV_{0}=2$ m$%
%TCIMACRO{\unit{eV}}%
%BeginExpansion
\operatorname{eV}%
%EndExpansion
$.\begin{figure}[t]
\centering\includegraphics[width=8.6cm]{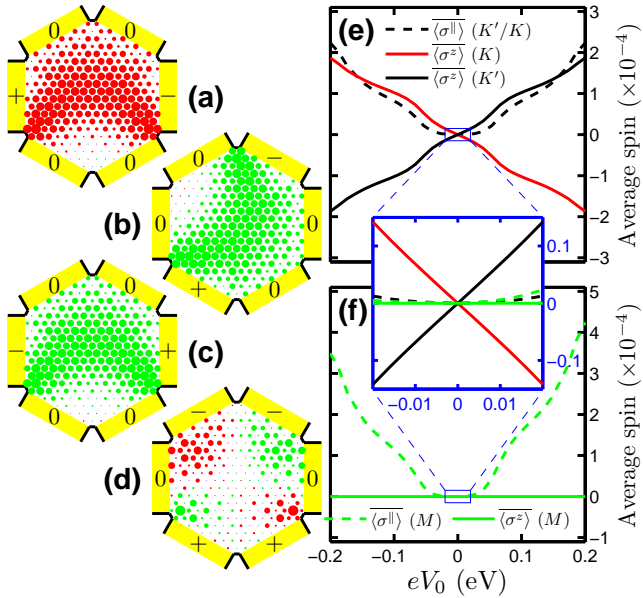}\caption{{}(Online
color) Surface spin polarization in a 6-terminal bilayer honeycomb channel
with (a) $\bar{K}^{\prime}$biasing, (b) and (c) $\bar{K}$ biasing, and (d)
$\bar{M}$ biasing. Red/gray (green/light gray) dots represent a $\langle
\sigma^{z}\rangle>0$ ($\langle\sigma^{z}\rangle<0$) local spin density with
the dot size proportional to $\left\vert \langle\sigma^{z}\rangle\right\vert
$. $\overline{\langle\sigma^{z}\rangle}$ and $\overline{\langle\sigma
^{\parallel}\rangle}$ as functions of the bias $eV_{0}$ are shown in (e) for
$\bar{K}$/$\bar{K}^{\prime}$ biasing and in (f) for $\bar{M}$ biasing.}%
\label{FIG4}%
\end{figure}

First we drive the electrons from left to right, corresponding to $\bar
{K}^{\prime}$. As expected, we have a positive average of surface spin
polarization $\overline{\langle\sigma^{z}\rangle}>0$, as shown in Fig.
\ref{FIG4}(a). When rotating the bias direction by $60^{\circ}$
counterclockwise, the surface polarization becomes $\overline{\langle
\sigma^{z}\rangle}<0$ as a consequence of Eq. (\ref{SK = -SK'}) [see Fig.
\ref{FIG4}(b)]. Reversing the bias of Fig. \ref{FIG4}(a) also switches
$\bar{K}^{\prime}$ states to $\bar{K}$, leading to $\overline{\langle
\sigma^{z}\rangle}<0,$ as shown in Fig. \ref{FIG4}(c). In Fig. \ref{FIG4}(d)
we drive the electrons from bottom to top, corresponding to the $\bar{M}$
direction. The spin density distribution becomes completely different and
satisfies the intrinsic spin-Hall symmetry, which yields $\overline
{\langle\sigma^{z}\rangle}=0$. Figures \ref{FIG4}(a) and \ref{FIG4}(c) imply
that the out-of-plane surface spin polarization can be flipped simply by
reversing the bias, which is also a direct consequence of time-reversal
operation. To show this electrical control of surface spin, we plot
$\overline{\langle\sigma^{z}\rangle}$ and $\overline{\langle\sigma_{\parallel
}\rangle}\equiv(\overline{\langle\sigma^{x}\rangle}^{2}+\overline
{\langle\sigma^{y}\rangle}^{2})^{1/2}$ as a function of the bias $eV_{0}$ in
Figs. \ref{FIG4}(e) and \ref{FIG4}(f) for $\bar{K}^{\prime}$/$\bar{K}$ and
$\bar{M}$ biasing, respectively. Within the low bias regime [see the inset
between Figs. \ref{FIG4}(e) and \ref{FIG4}(f)], $\overline{\langle\sigma
^{z}\rangle}$ for $\bar{K}^{\prime}$/$\bar{K}$ biasing grows with $eV_{0}$
linearly, while other components are either vanishing or relatively small.

In conclusion, we have presented a unified tight-binding description to
understand the Rashba effect in graphene, as well as bilayer surfaces and
surface alloys of the honeycomb structure. Our results explain the recently
observed abrupt upstanding Rashba spin in Tl/Si(111)-$\left(  1\times1\right)
$ surface alloy around $\bar{K}$, and predict an electrically reversible
out-of-plane surface spin polarization, which may serve as a storage mechanism
for future spintronic devices.

Financial support of the Republic of China National Science Council (Grant No.
NSC 98-2112-M-002-012-MY3) is gratefully acknowledged.

\bibliographystyle{apsrev}
\bibliography{mhl2}

\end{document}